\documentclass[%
 reprint,
 amsmath,amssymb,
 aps,
]{revtex4-1}

\usepackage{graphicx}
\usepackage{dcolumn}
\usepackage{bm}

\begin{document}

\preprint{APS/123-QED}

\title{Schwinger-Dyson Equation Method to Calculate Total Energy in Periodic Anderson Model with Electron-Phonon Interactions}
\author{Enzhi Li  \email{enzhililsu@gmail.com}}
\affiliation {Suning R\&D Center, Palo Alto, USA}

\date{\today}

\begin{abstract}
We have recently employed periodic Anderson model with electron-phonon interactions to describe Cerium volume collapse ($\gamma \rightarrow \alpha$ transition) under pressure. To describe the volume collapse transition in Cerium, we have tried to plot the pressure versus volume curves and see when a kink structure emerges. One way to obtain the pressure versus volume curve is to calculate the total energy of the system at different temperatures. In order to solve the periodic Anderson model with electron-phonon interactions, we have used the continuous time quantum Monte Carlo algorithm by integrating out the phonons to obtain a retarded electron-electron interaction. Monte Carlo simulation results give us the model's electronic Green function and self energy, from which we can calculate the system's total energy. A well known formula for calculating total energy through the knowledge of electron's self energy and Green function is derived from the model's Hamiltonian, which is not available for our case. Here, we have devised a new method to derive the total energy formula purely from a path integral description of the system, without resort to its Hamiltonian formulation. Since not all systems can be described by a Hamiltonian, thus our method has a wider applicability. It is noteworthy that the total energy formula that we derived here takes an identical form with that obtained from a model's Hamiltonian. 
\end{abstract}
\pacs{}
\maketitle

\section{Introduction}
Recently, we employ periodic Anderson model with electron-phonon interactions to study the behavior of $4f$ electrons in Cerium to gain further understanding about the volume collapse that Cerium experiences at high pressure\cite{Peng_2013}. We have introduced electron-phonon interactions to the periodic Anderson model to capture the influence of lattice dynamics upon the Cerium volume collapse. It is argued that the Cerium volume collapse should be due to the emergence of a first order phase transition. A good method to characterize this first order phase transition is to obtain pressure versus volume curves at different temperatures and see when the pressure versus volume curve develops a kink\cite{PhysRevB.99.155147}. We can calculate the pressure by taking partial derivative of free energy with respect to volume, whereas free energy can be calculated once we have obtained energy as a function of temperature\cite{PhysRevB.80.140505}. Thus, a central issue we encountered is to calculate the total energy for the periodic Anderson model with electron-phonon interactions. It is well know that once we have a model's self energy and Green function, we can calculate its total energy via this formula(see, for example, Ref. \cite{fetter2003quantum}): 
\begin{align}
\label{energy_formula}
E &= \frac{1}{\beta}\sum_{i\omega_n}\sum_{\boldsymbol{k}, \sigma}\epsilon_{\boldsymbol{k}}G_{\boldsymbol{k}, \sigma}(i\omega_n) \\\nonumber
& + \frac{1}{\beta}\sum_{i\omega_n}\sum_{\boldsymbol{k}, \sigma}\Sigma_{\boldsymbol{k}, \sigma}(i\omega_n)G_{\boldsymbol{k}, \sigma}(i\omega_n), 
\end{align}
where $\epsilon_{\boldsymbol{k}}$ is electron's dispersion relation in a lattice, $\Sigma_{\boldsymbol{k}, \sigma}(i\omega_n)$ is electron's self energy, and $G_{\boldsymbol{k}, \sigma}(i\omega_n)$ is electron's Green function, with $\omega_{n} = (2n + 1)\pi/\beta$ being the Matsubara frequency. 
This formula is a natural corollary of equation of motion for electron's Green function, which can be conveniently derived once we have the model's Hamiltonian. However, not all physical systems are susceptible to a Hamiltonian formulation, and a notable family of such systems are the ones that are derived from the principle of least action\cite{feynman2005feynman}. It is universally known that the quantum analogue of the principle of least action is Feynman's path integral method. Therefore, for the systems that can only be properly described by a path integral, the validity of Eq. [\ref{energy_formula}] which is derived with the implicit assumption for the existence of a Hamiltonian is at stake, and a justification or refutation of that formula is on its way. 

For our case, in order to solve the periodic Anderson model with electron-phonon interactions, i.e., obtain the electron's Green function and self energy, we have employed the diagrammatic Monte Carlo method as detailed in Ref. \cite{CTQMC_2007}, where the authors have integrated out the phonons to obtain a time-retarded electron-electron interaction. Once the phonons have been integrated out, we are left with an effective action, from which no explicit Hamiltonian can be obtained. Because a Hamiltonian is directly related with the total energy of a system, it becomes tricky to calculate a system's total energy without an explicit Hamiltonian formulation. It it however noteworthy that not all systems with retarded interactions are bereft of a Hamiltonian formulation, see, for example, Ref. \cite{PhysRevB.96.075155}, where the authors can still obtain the total energy of a system with retarded interactions using operator method due to the availability of a Hamiltonian formulation for their system. In our case for which no explicit Hamiltonian formulation is available, we are going to devise a new method to calculate the total energy using effective action, rather than Hamiltonian, as our starting point. We will show that the formula (Eq. [\ref{energy_formula}]) for calculating the total energy which was originally derived from the Hamiltonian is still applicable here when the system can only be described by an effective action, although the mathematical details for the formula derivation are completely different. 

The problem of dealing quantum mechanically with a system that is derived from the principle of least action is encountered by Feynman, who designed the path integral approach to quantum mechanics just to address such an issue\cite{RevModPhys.20.367, feynman2005feynman, feynman1965quantum}. By integrating out the phonons in our model, we also encounter a similar situation, where we are left with an effective system with no Hamiltonian formulation. In order to derive the total energy formula for such a system, we need to employ the Schwinger-Dyson equation method, which furnishes us with an equation motion for Green function which is purely derived from path integrals. Once we have the equation of motion for electron's Green function, we can easily calculate the total energy of the system. Therefore, the central topic of this paper is the equation of motion for Green function, which we shall derive using several different methods. The equivalence of these methods is also a major topic of this paper.  

The organization of the paper is as follows. In section \ref{toy_holstein_model}, we will show the equivalence of Schwinger-Dyson equation method and the operator method in the derivation of equations of motion in a toy model that nevertheless captures the essence of electron-phonon interactions. In this section, we will derive the equation of motion using three methods: the operator method, the path integral method with phonons present and the path integral method with phonons integrated out. All of these methods yield identical results. Having established the validity of Schwinger-Dyson equation method in this toy model, we will continue to show in section \ref{pam_model} that we can derive the equation of motion, and thus the total energy, of periodic Anderson model with electron-phonon interactions (PAM + phonons). Since we have used the continuous time quantum Monte Carlo algorithm to simulate PAM + phonons model, and during the Monte Carlo simulation process it is advantageous to integrate out phonons to obtain an effective model that contains retarded electron-electron interactions, we thus need to calculate the total energy of a system without an explicit Hamiltonian formulation. It is this situation that compels us to develop the Schwinger-Dyson equation \cite{schwinger1951green, peskin2018introduction} method to derive the equation of motion as described in this paper. Finally, we make a conclusion in section \ref{conclusion}. In the appendix, we give more details about how to integrate out phonons in periodic Anderson with electron-phonon interactions. 

\section{Schwinger-Dyson equation method to derive equation of motion for a simplified Holstein model}
\label{toy_holstein_model}
In this section, we are going to present the derivation of the equation of motion for Green function in a simplified Holstein model using three different methods: the operator method, the path integral method with phonons present, and the path integral method with phonons integrated out. The operator method, although most straightforward, is restricted to the case where we have a Hamiltonian formulation of the system, a formulation that is not always available. We will next show that we can get the same results as that from operator method using path integral formulation. Once we have formulated our system using path integrals, we can integrate out the phonons and obtain an effective action that only involves electrons. We shall demonstrate that the equation of motion of the electron's Green function, and thus the total energy of the system, remains invariant regardless whether the phonons are retained or integrated out. In the Continuous time quantum Monte Carlo simulation process, it is advantageous to integrate out the phonons and obtain a retarded potential between electrons. The results obtained in this section indicate that we can use the same equation to calculate the total energy even if the phonons are integrated out and a Hamiltonian formulation of the system is unavailable. It is noteworthy that the Schwinger-Dyson equation method applies to any model that admits a path integral formulation, especially models with electron-phonon interactions. 

\subsection{Model description}
Our toy model can be described by this Hamiltonian:  
\begin{eqnarray}
\hat{H} = \epsilon\hat{c}^{\dagger}\hat{c} + \omega\hat{a}^{\dagger}\hat{a} + g\hat{c}^{\dagger}\hat{c}(\hat{a} + \hat{a}^{\dagger})
\end{eqnarray}
This model contains one spinless electron with energy $\epsilon$, one dispersionless phonon with energy $\omega$, and an interaction term between the electron and the phonon. 
The partition function for this Hamiltonian is 
\begin{eqnarray}
Z &=& \text{Tr} e^{-\beta\hat{H}} \\\nonumber
&=& \int \mathcal{D}[\bar{c}, c]\mathcal{D}[\bar{a}, a] e^{-S}, 
\end{eqnarray}
where the action of the system is\cite{coleman2015introduction}
\begin{eqnarray}
S = \int_{0}^{\beta} d\tau \Big( \bar{c}(\partial_{\tau} + \epsilon)c + \bar{a}(\partial_{\tau} + \omega)a + g\bar{c}c(a + \bar{a}) \Big)
\end{eqnarray}
Since the complex variables $\bar{a}, a$ are quadratic in the exponent, we can perform the Gaussian integral in a closed form as 
\begin{align}
Z &= Z_0 Z_{eff} \\\nonumber
&= \frac{1}{1 - e^{-\beta\omega}} \int \mathcal{D}[\bar{c}, c] e^{-S_{eff}}, 
\end{align}
where the effective action $S_{eff}$ is 
\begin{widetext}
\begin{eqnarray}
\label{Seff_toy_model}
S_{eff} = \int_{0}^{\beta} d\tau \bar{c}(\tau) (\partial_{\tau} + \epsilon)c(\tau) + \frac{g^2}{2} \iint d\tau d\tau^{\prime} \bar{c}(\tau) c(\tau) D_{0}(\tau - \tau^{\prime}, \omega) \bar{c}(\tau^{\prime})c(\tau^{\prime})
\end{eqnarray}
\end{widetext}
In the above equation for $S_{eff}$, we have defined the bare phonon propagator: 
\begin{eqnarray}
 D_{0}(\tau, \omega) = -\frac{e^{\omega|\tau |}}{e^{\beta\omega} - 1} - \frac{e^{-\omega|\tau |}}{1 - e^{-\beta\omega}}, -\beta < \tau < \beta
\end{eqnarray}
Electron's Green function for Hamiltonian that does not depend on $\tau$ explicitly is defined as (using operator notation)
\begin{align}
G(\tau^{\prime} - \tau) = -\langle T\hat{c}(\tau^{\prime})\hat{c}^{\dagger}(\tau) \rangle
\label{Green_operator}
\end{align}
It can be also be equivalently defined as (using path integral formulation)
\begin{align}
G(\tau^{\prime} - \tau) = -\frac{1}{Z}\int \mathcal{D}[\bar{c}, c] \mathcal{D}[\bar{a}, a] c(\tau^{\prime})\bar{c}(\tau) e^{-S}
\label{Green_path_integral}
\end{align}
Now we are going to derive the equation of motion for the electron's Green function using both operator notation and path integral formulation, and show that these methods yield identical result. 

\subsection{Operator method to derive equation of motion for Green function}
Derivation of the equation of motion for Green function using operator notation method is pretty straightforward. We only need to apply time derivative operator $\frac{\partial}{\partial\tau}$ to Equ. [\ref{Green_operator}] and obtain this result:  
\begin{widetext}
\begin{eqnarray}
\frac{\partial G(\tau^{\prime}, \tau)}{\partial \tau^{\prime}} = -\delta(\tau^{\prime} - \tau) + \epsilon\langle T \hat{c}(\tau^{\prime})\hat{c}^{\dagger}(\tau)\rangle + g\langle T\hat{c}(\tau^{\prime})\hat{\phi}(\tau^{\prime})\hat{c}^{\dagger}(\tau)\rangle
\end{eqnarray}
That is, 
\begin{eqnarray}
(\partial_{\tau^{\prime}} + \epsilon)G(\tau^{\prime}, \tau) = -\delta(\tau^{\prime} - \tau) + g\langle T\hat{c}(\tau^{\prime})\hat{\phi}(\tau^{\prime})\hat{c}^{\dagger}(\tau)\rangle
\label{eom_1}
\end{eqnarray}
\end{widetext}
Here, we have introduced the phonon displacement operator $\hat{\phi} = \hat{a} + \hat{a}^{\dagger}$. This method is easy to understand and straightforward to follow. However, in order to use this method, we must have a Hamiltonian formulation of the system, which is not always available. Feynman originally designed path integral formulation of quantum mechanics just to cope with the situation where the system is derived from the principle of least action and cannot be described using a Hamiltonian. We also encounter the same situation while employing CT-QMC to simulate models with electron-phonon interactions. Next, we will show that we can equally well derive the equation of motion for Green function even with the absence of an explicit Hamiltonian. The way to circumvent Hamiltonian is to use the path integral method, which we will describe in detail next. 

\subsection{Path integral method to derive equation of motion for Green function: with phonons present}
As we can see, the equation of motion for Green's function is easy to obtain as long as we have an operator formulation of the system. We can equivalently derive the equation of motion using path integral formulation. For this, we are to employ the Schwinger-Dyson equation. Schwinger-Dyson equation is not well known in condensed matter community, and thus we shall give this equation a brief derivation that should be easy to follow for anyone who is familiar with variational calculus. First, consider the following infinitesimal translation of the Grassmann field variables: 
\begin{eqnarray}
&& c \rightarrow c^{\prime}  = c + \delta c, \\\nonumber
&& \bar{c} \rightarrow \bar{c}^{\prime} = \bar{c} + \delta \bar{c}, 
\end{eqnarray}
\begin{widetext}
By definition, the following quantity should be invariant before and after the field translation: 
\begin{eqnarray}
\int \mathcal{D}[\bar{c}, c]\mathcal{D}[\bar{a}, a] \bar{c}(\tau)e^{-S[\bar{c}, c; \bar{a}, a]} = \int \mathcal{D}[\bar{c}^{\prime}, c^{\prime}] \mathcal{D}[\bar{a}, a] \bar{c}^{\prime}(\tau) e^{-S[\bar{c}^{\prime}, c^{\prime}; \bar{a}, a]}
\label{invariant_quantity}
\end{eqnarray}
However, the action $S$ does change under this translation. In the notation of the translated field variables, the action is 
\begin{align}
S[\bar{c}^{\prime}, c^{\prime}; \bar{a}, a] &= S[\bar{c}, c; \bar{a}, a] \\\nonumber
&+\int_{0}^{\beta}d\tau \Big( \delta\bar{c}(\partial_{\tau} + \epsilon)c + \bar{c}(\partial_{\tau} + \epsilon)\delta c + g\delta\bar{c}c(a + \bar{a}) + g\bar{c}\delta c (a + \bar{a}) \Big) \\\nonumber
&=: S[\bar{c}, c; \bar{a}, a] + \delta S
\end{align}
Taking advantage of the fact that $\mathcal{D}[\bar{c}, c] = \mathcal{D}[\bar{c}^{\prime}, c^{\prime}]$ since the Jacobian determinant for translation is unity, we can express Eq. [\ref{invariant_quantity}] equivalently as 
\begin{align}
\int \mathcal{D}[\bar{c}, c]\mathcal{D}[\bar{a}, a] \bar{c}(\tau)e^{-S[\bar{c}, c; \bar{a}, a]}  &= \int \mathcal{D}[\bar{c}, c]\mathcal{D}[\bar{a}, a] (\bar{c}(\tau) + \delta\bar{c}(\tau)) e^{-S[\bar{c}, c; \bar{a}, a] - \delta S}\\\nonumber
&= \int \mathcal{D}[\bar{c}, c]\mathcal{D}[\bar{a}, a] (\bar{c}(\tau) + \delta\bar{c}(\tau)) e^{-S}(1 - \delta S)\\\nonumber
&= \int \mathcal{D}[\bar{c}, c]\mathcal{D}[\bar{a}, a] \bar{c}(\tau)e^{-S[\bar{c}, c; \bar{a}, a]} + \int \mathcal{D}[\bar{c} ,c]\mathcal{D}[\bar{a}, a] e^{-S}(\delta\bar{c}(\tau) - \bar{c}(\tau)\delta S)
\end{align}
After cancellation of equal terms in the above equation, we finally arrive at the Schwinger-Dyson equation: 
\begin{eqnarray}
0 = \int \mathcal{D}[\bar{c} ,c]\mathcal{D}[\bar{a}, a] e^{-S}(\delta\bar{c}(\tau) - \bar{c}(\tau)\delta S)
\label{Schwinger_Dyson_equation}
\end{eqnarray}
This equation should hold for any field variations $\delta c, \delta\bar{c}$. In order to get the equation of motion for Green function, we set $\delta c = 0$, and the integrand in the above equation simplifies into this form: 
\begin{eqnarray}
&&\delta\bar{c}(\tau) - \bar{c}(\tau)\delta S \\\nonumber
&=& \delta\bar{c}(\tau) - \bar{c}(\tau)\int_{0}^{\beta}d\tau^{\prime} (\delta\bar{c}(\partial_{\tau^{\prime}} + \epsilon)c + g\delta\bar{c}c(a + \bar{a}))\\\nonumber
&=& \int_{0}^{\beta}d\tau^{\prime}\Big(\delta(\tau^{\prime} - \tau)\delta\bar{c}(\tau^{\prime}) - \delta\bar{c}(\tau^{\prime})(\partial_{\tau^{\prime}} + \epsilon)c(\tau^{\prime})\bar{c}(\tau) - g\delta\bar{c}(\tau^{\prime})c(\tau^{\prime})(a + \bar{a})\bar{c}(\tau)\Big)
\end{eqnarray}
Plug this back into Equ. [\ref{Schwinger_Dyson_equation}], we have 
\begin{eqnarray}
&&\int_{0}^{\beta}d\tau^{\prime}\Bigg[\delta(\tau - \tau^{\prime}) - (\partial_{\tau^{\prime}} + \epsilon) \frac{1}{Z}\int\mathcal{D}[\bar{c}, c]\mathcal{D}[\bar{a}, a] e^{-S} c(\tau^{\prime}) \bar{c}(\tau) \\\nonumber
&-& g\frac{1}{Z}\int \mathcal{D}[\bar{c},c]\mathcal{D}[\bar{a}, a]e^{-S}c(\tau^{\prime})\phi(\tau^{\prime})\bar{c}(\tau)\Bigg]\delta\bar{c}(\tau^{\prime}) = 0
\end{eqnarray}
Since this equation holds for any $\delta\bar{c}$, we thus, from the definition of Green function in path integral formulation, have 
\begin{eqnarray}
(\partial_{\tau^{\prime}} + \epsilon)G(\tau^{\prime}, \tau) - g\langle T\hat{c}(\tau^{\prime})\hat{\phi}(\tau^{\prime})\hat{c}^{\dagger}(\tau) \rangle = -\delta(\tau - \tau^{\prime})
\label{eom_2}
\end{eqnarray}
This is identical to the equation of motion that was derived using operator notation. 
\end{widetext}

\subsection{Path integral method to derive equation of motion for Green function: with phonons integrated out}
The advantage of Schwinger-Dyson equation method for derivation of equation of motion for Green function is that it applies to any system that admits a path integral formulation. Following Feynman, we can reformulate a large family of classical Hamiltonian systems into its quantum counterpart using path integrals, however, as Feynman noted in his thesis, there are systems that can only be represented using classical action for which no Hamiltonian expression can be found. In this case, the Schwinger-Dyson equation is the only method available for the derivation of equation of motion. One such case when the system can only be represented in path integrals is the Holstein model with phonons integrated out. This is the standard approach to simulate electron-phonon interactions using quantum Monte Carlo. In order to study the thermodynamical properties of the system, we need to known the model's entropy and free energy, and the numerical computation of these two quantities requires the calculation of energy. One method to derive model's energy is to use the equation of motion for Green function. Thus, it is an imperative task to know how to derive the equation of motion for Green function without an explicit presence of Hamiltonian that we usually take for granted. For our simplified Holstein model, the action is  
\begin{eqnarray}
S = \int_{0}^{\beta}d\tau \bar{c}(\tau)(\partial_{\tau} + \epsilon)c(\tau) + S_{\phi}, 
\end{eqnarray}
where $S_{\phi}$ is the part of the action that contains phonons: 
\begin{eqnarray}
S_{\phi} = \int_{0}^{\beta}d\tau \Big(\bar{a}(\partial_{\tau} + \omega)a + g\bar{c}c(a + \bar{a})\Big)
\end{eqnarray}
By integrating out phonons, we can recast the partition function into this form:  
\begin{align}
Z &= \int \mathcal{D}[\bar{c}, c]\mathcal{D}[\bar{a}, a] e^{-S} \\\nonumber
&= Z_{\phi}\int \mathcal{D}[\bar{c}, c] e^{-S_{eff}}
\end{align}
Here, $Z_{\phi} = \Big(1- e^{-\beta\omega}\Big)^{-1}$ is the partition function for bare phonons, and $S_{eff}$ whose explicit form is already given in Eq. [\ref{Seff_toy_model}] is the effective action obtained by integrating out the phonons.
Similarly, define a quantity that is invariant under the transformation $c\rightarrow c^{\prime} = c + \delta c, \bar{c}\rightarrow\bar{c}^{\prime} + \delta \bar{c}^{\prime}$, that is,
\begin{widetext}
\begin{eqnarray}
\int \mathcal{D}[\bar{c},c] e^{-S_{eff}[\bar{c} ,c]} \bar{c}(\tau) = \int\mathcal{D}[\bar{c}^{\prime}, c^{\prime}] e^{-S_{eff}[\bar{c}^{\prime}, c^{\prime}]}\bar{c}^{\prime}(\tau)
\end{eqnarray}
Using the techniques that we employed for the derivation of Eq. [\ref{Schwinger_Dyson_equation}], we arrive at the Schwinger-Dyson equation for this model as  
\begin{eqnarray}
0 = \int\mathcal{D}[\bar{c},c ]e^{-S_{eff}[\bar{c},c]}(-\delta S_{eff}\bar{c}(\tau) + \delta\bar{c}(\tau))
\label{Schwinger_Dyson}
\end{eqnarray}
If we set $\delta c = 0$ and notice that $D^{0}(\tau_1 - \tau_2) = D^{0}(\tau_2 - \tau_1)$, then we obtain the variation of the action as
\begin{eqnarray}
\delta S_{eff} = \int_{0}^{\beta}d\tau\delta\bar{c}(\partial_{\tau} + \epsilon)c + g^2\int_{0}^{\beta}d\tau_1\int_{0}^{\beta}d\tau_2\delta\bar{c}(\tau_1)c(\tau_1)D^{0}(\tau_1 - \tau_2)\bar{c}(\tau_2)c(\tau_2)
\end{eqnarray}
Thus the integrand in Equ. [\ref{Schwinger_Dyson}] is 
\begin{eqnarray}
&&\delta\bar{c}(\tau) - \delta S_{eff}\bar{c}(\tau) \\\nonumber
&=& \int_{0}^{\beta}d\tau^{\prime}\delta\bar{c}(\tau^{\prime})\Bigg[ \delta(\tau - \tau^{\prime}) - (\partial_{\tau^{\prime}} + \epsilon)c(\tau^{\prime})\bar{c}(\tau) - g^2c(\tau^{\prime})\bar{c}(\tau)\int_{0}^{\beta}d\tau_2D^{0}(\tau^{\prime} - \tau_2)\bar{c}(\tau_2)c(\tau_2)\Bigg]
\end{eqnarray}
We demand that Eq. [\ref{Schwinger_Dyson}] should hold for any field variation $\delta\bar{c}$. From the principles of variational calculus, we obtain the equation of motion for Green function: 
\begin{align}
\label{eom_3}
&(\partial_{\tau^{\prime}} + \epsilon)G(\tau^{\prime}, \tau) \\\nonumber
&- g^{2}\frac{1}{Z}\int\mathcal{D}[\bar{c},c]e^{-S_{eff}}\Bigg[c(\tau^{\prime})\bar{c}(\tau)\int_{0}^{\beta} d\tau_2 D^{0}(\tau^{\prime} - \tau_2)\bar{c}(\tau_2)c(\tau_2)\Bigg] = -\delta(\tau^{\prime}  - \tau)
\end{align}
The equivalence of Eq. [\ref{eom_1}], Eq. [\ref{eom_2}] and Eq. [\ref{eom_3}] is easy to see. In Eq. [\ref{eom_3}], if we set $\tau = 0, \tau^{\prime} = 0^{-}$, then we have the expectation value of the interaction energy as 
\begin{align}
\label{energy_path_integral}
&g^{2}\frac{1}{Z}\int\mathcal{D}[\bar{c}, c] e^{-S_{eff}}\Bigg[\bar{c}(0)c(0)\int_{0}^{\beta}d\tau_2 D^{0}(\tau_2)\bar{c}(\tau_2)c(\tau_2)\Bigg] \\\nonumber
=& -(\partial_{\tau^{\prime}} + \epsilon)G(\tau^{\prime})\Big\vert_{\tau^{\prime} = 0^{-}} - \delta(0^{-})\\\nonumber
=& \frac{1}{\beta}\sum_{i\omega_n}\Sigma(i\omega_n)G(i\omega_n)e^{i\omega_n 0^{+}}
\end{align}
\end{widetext}
Similarly, for Eq. [\ref{eom_1}] and [\ref{eom_2}], we can also set $\tau^{\prime} = 0^{-}, \tau = 0$ to obtain the electron-phonon interaction energy as 
\begin{align}
\label{energy_operator}
g\langle \hat{c}^{\dagger} \hat{c} \hat{\phi} \rangle = \frac{1}{\beta}\sum_{i\omega_n}\Sigma(i\omega_n) G(i\omega_n) e^{i\omega_n 0^{+}}
\end{align}

The equivalence of Eq. [\ref{energy_path_integral}] and Eq. [\ref{energy_operator}] is obvious. Note that the total energy can be calculated by adding electronic kinetic energy which is trivial to obtain to the electron-phonon interaction energy. Thus far, we have derived the total energy formula via three distinct methods: the operator method, the path integral method with phonon present, and the path integral method with phonons integrated out. All three methods yield identical result. 

In this section, we have demonstrated the applicability of Schwinger-Dyson equation method for the derivation of total energy formula for a toy Holstein model. In the next section, we will employ Schwinger-Dyson equation to derive the total energy formula for periodic Anderson model with electron-phonon interactions following the procedures that are delineated in this section. 

\section{Schwinger-Dyson method to derive equation of motion for periodic Anderson model with electron-phonon interactions }
\label{pam_model}

\subsection{Periodic Anderson model with electron-phonon interactions: A Brief description}
Our model Hamiltonian for periodic Anderson model with electron-phonon interaction is 
\begin{align}
\hat{H} &= \hat{H}_{0} + \hat{H}_{I} 
\label{eq:Hamiltonian}\\\nonumber
\hat{H}_{0} &= -t\sum_{\langle i, j \rangle, \sigma} (c_{i,\sigma}^{\dagger} c_{j,\sigma} + c_{j,\sigma}^{\dagger} c_{i,\sigma}) + \epsilon_{f} \sum_{i, \sigma} f_{i, \sigma}^{\dagger}f_{i, \sigma} \\\nonumber
& + V \sum_{i,\sigma} (c_{i,\sigma}^{\dagger}f_{i,\sigma} + f_{i,\sigma}^{\dagger}c_{i,\sigma}) + \sum_{i}\omega_{0}\hat{a}_{i}^{\dagger}\hat{a}_{i} \\\nonumber
\hat{H}_{I} &= U\sum_{i} n_{i,\uparrow}^{f}n_{i,\downarrow}^{f} + g \sum_{i, \sigma}(\hat{a}_{i}^{\dagger} + \hat{a}_{i})\hat{c}_{i, \sigma}^{\dagger}\hat{c}_{i, \sigma}
\end{align}
Here, $c_{i, \sigma}^{\dagger}, c_{i, \sigma} (f_{i, \sigma}^{\dagger}, f_{i, \sigma})$ creates and destroys a $c(f)$ electron of spin $\sigma$ at lattice site $i$, respectively.  $\hat{a}_{i}$ ($\hat{a}^{\dagger}_{i}$) is the annihilation (creation) operator for a phonon located at lattice site $i$. $U$ is the Hubbard repulsion between localized $f$-electrons, and $V$ characterizes the hybridization between $c$ and $f$ electrons. 
This Hamiltonian in momentum space is 
\begin{align}
\hat{H} &= \hat{H}_0 + \hat{H}_I\\ \nonumber
\hat{H}_0 &= \sum_{\boldsymbol{k}, \sigma} \left(
\begin{array}{cc}
 \hat{c}_{\boldsymbol{k}\sigma}^{\dagger} & \hat{f}_{\boldsymbol{k}\sigma}^{\dagger} \\
\end{array}
\right).\left(
\begin{array}{cc}
 \epsilon_{\boldsymbol{k}} & V \\
 V & \epsilon_{f} \\
\end{array}
\right).\left(
\begin{array}{c}
 \hat{c}_{\boldsymbol{k}\sigma} \\
 \hat{f}_{\boldsymbol{k}\sigma} \\
\end{array}
\right)
 + \sum_{\boldsymbol{k}}\omega_0 \hat{a}^{\dagger}_{\boldsymbol{k}}\hat{a}_{\boldsymbol{k}}\\ \nonumber
 \hat{H}_{I} &= \sum_{\boldsymbol{k}\boldsymbol{q}\sigma}g\hat{c}_{\boldsymbol{k} + \boldsymbol{q}, \sigma}^{\dagger}\hat{c}_{\boldsymbol{k}\sigma}\hat{\phi}_{\boldsymbol{q}} + U\sum_{\boldsymbol{k}\boldsymbol{p}\boldsymbol{q}}\hat{f}^{\dagger}_{\boldsymbol{p}\uparrow}
 \hat{f}_{\boldsymbol{k}\uparrow}\hat{f}^{\dagger}_{\boldsymbol{q}\downarrow}\hat{f}_{\boldsymbol{p} + \boldsymbol{q} - \boldsymbol{k}, \downarrow}\\\nonumber
 \hat{\phi}_{\boldsymbol{q}} &= \hat{a}_{\boldsymbol{q}} + \hat{a}^{\dagger}_{-\boldsymbol{q}}
\end{align}
We are going to use the path integral method to integrate out the phonons and obtain an effective action that describes the retarded density-density interactions between electrons. With path integrals, the partition function for this Hamiltonian can be written as 
\begin{eqnarray}
Z = \int \mathcal{D}[\bar{c}, c]\mathcal{D}[\bar{f}, f]\mathcal{D}[\bar{a}, a] e^{-S}
\end{eqnarray}
Integrating out the phonons, we can factorize the partition function into the form $Z = Z_{0} Z_{eff}$, with $Z_{0} = (1 - e^{-\beta\omega})^{-1}$, and $Z_{eff} = \int\mathcal{D}[\bar{c}, c] \mathcal{D}[\bar{f}, c] e^{-S_{eff}}$. The effective action is 
\begin{widetext}
\begin{align}
\label{S_eff_PAM}
S_{eff} &= \int_{0}^{\beta}d\tau\sum_{\boldsymbol{k}, \sigma} \left(
\begin{array}{cc}
 \bar{c}_{\boldsymbol{k}\sigma} & \bar{f}_{\boldsymbol{k}\sigma} \\
\end{array}
\right).\left(
\begin{array}{cc}
 \partial_{\tau} + \epsilon_{\boldsymbol{k}} & V \\
 V & \partial_{\tau} + \epsilon_{f} \\
\end{array}
\right).\left(
\begin{array}{c}
 c_{\boldsymbol{k}\sigma} \\
f_{\boldsymbol{k}\sigma} \\
\end{array}
\right) \\\nonumber
&+ \frac{1}{2}g^2 \sum_{\boldsymbol{q}}\int_{0}^{\beta}d\tau_1\int_{0}^{\beta}d\tau_2\sum_{\boldsymbol{k}, \sigma}\bar{c}_{\boldsymbol{k},\sigma}(\tau_1)c_{\boldsymbol{k} - \boldsymbol{q}, \sigma}(\tau_1)D^{0}(\tau_1 - \tau_2)\sum_{\boldsymbol{p},s}\bar{c}_{\boldsymbol{p} - \boldsymbol{q}, s}(\tau_2)c_{\boldsymbol{p},s}(\tau_2) \\\nonumber
&- U\int_{0}^{\beta}d\tau \sum_{\boldsymbol{k}\boldsymbol{p}\boldsymbol{q}} \bar{f}_{\boldsymbol{p}\uparrow}\bar{f}_{\boldsymbol{q}\downarrow}f_{\boldsymbol{k}\uparrow}f_{\boldsymbol{p} + \boldsymbol{q} - \boldsymbol{k}, \downarrow}
\end{align}
\end{widetext}
See the appendix for a detailed derivation of the effective action. 
With the effective action, we can employ the Schwinger-Dyson equation to derive the equation of motion for electron's Green function, and further calculate the total energy of the Hamiltonian. Next, we will show how to derive the equation of motion for Green function.

\subsection{Equation of motion}
Now that we have already obtained the effective action for our model, we shall begin to derive the equation of motion for Green function using Schwinger-Dyson equation method. 
Consider the variations of field variables, 
\begin{widetext}
\begin{eqnarray}
&& c_{k,\sigma}(\tau) \rightarrow c^{\prime}_{k,\sigma}(\tau) = c_{k,\sigma}(\tau) + \delta c_{k,\sigma}(\tau) \\\nonumber
&& \bar{c}_{k,\sigma}(\tau) \rightarrow \bar{c}^{\prime}_{k,\sigma}(\tau) = \bar{c}_{k,\sigma}(\tau) + \delta \bar{c}_{k,\sigma}(\tau) \\\nonumber
&& f_{k,\sigma}(\tau) \rightarrow f^{\prime}_{k,\sigma}(\tau) = f_{k,\sigma}(\tau) + \delta f_{k,\sigma}(\tau)\\\nonumber
&& \bar{f}_{k,\sigma}(\tau) \rightarrow \bar{f}^{\prime}_{k,\sigma}(\tau) = \bar{f}_{k,\sigma}(\tau) + \delta \bar{f}_{k,\sigma}(\tau) 
\end{eqnarray}
With this transformation, the effective action becomes $S_{eff}[\bar{c}, c; \bar{f}, f] \rightarrow S_{eff}[\bar{c}^{\prime}, c^{\prime}; \bar{f}^{\prime}, f^{\prime}] = S_{eff}[\bar{c}, c; \bar{f}, f] + \delta S_{eff}$. Define a quantity that is invariant under this transformation, 
\begin{eqnarray}
&&\int \mathcal{D}[\bar{c}, c]\mathcal{D}[\bar{f}, f] e^{-S_{eff}[\bar{c}, c; \bar{f}, f]} \begin{pmatrix} \bar{c}_{k,\sigma}(\tau) && \bar{f}_{k,\sigma}(\tau)\end{pmatrix} \\\nonumber
&=& \int \mathcal{D}[\bar{c}^{\prime}, c^{\prime}]\mathcal{D}[\bar{f}^{\prime}, f^{\prime}] e^{-S_{eff}[\bar{c}^{\prime}, c^{\prime}; \bar{f}^{\prime}, f^{\prime}]} \begin{pmatrix}\bar{c}^{\prime}_{k,\sigma}(\tau) &&  \bar{f}^{\prime}_{k,\sigma}(\tau \end{pmatrix} \\\nonumber
&=& \int\mathcal{D}[\bar{c},c]\mathcal{D}[\bar{f}, f] e^{-S_{eff}[\bar{c}, c; \bar{f}, f] - \delta S_{eff}}\begin{pmatrix} \bar{c}_{k,\sigma}(\tau) + \delta \bar{c}_{k,\sigma}(\tau) &&  \bar{f}_{k,\sigma}(\tau) + \delta \bar{f}_{k,\sigma}(\tau) \end{pmatrix}
\end{eqnarray}
After expanding to first order and cancelling identical terms on both sides, we have 
\begin{eqnarray}
\label{pam_Schwinger_Dyson_equation}
0 = \int\mathcal{D}[\bar{c},c]\mathcal{D}[\bar{f},f] e^{-S_{eff}}\begin{pmatrix} \delta \bar{c}_{k,\sigma}(\tau) - \delta S_{eff}\bar{c}_{k,\sigma}(\tau) &&  \delta \bar{f}_{k,\sigma}(\tau) - \delta S_{eff}\bar{f}_{k,\sigma}(\tau) \end{pmatrix}
\end{eqnarray}
Here, the variation of action is 
\begin{eqnarray}
\delta S_{eff} &=& \int_{0}^{\beta}d\tau \sum_{\boldsymbol{k}, \sigma} \left(
\begin{array}{cc}
 \delta \bar{c}_{\boldsymbol{k}\sigma} & \delta \bar{f}_{\boldsymbol{k}\sigma} \\
\end{array}
\right).\left(
\begin{array}{cc}
 \partial_{\tau} + \epsilon_{\boldsymbol{k}} & V \\
 V & \partial_{\tau} + \epsilon_{f} \\
\end{array}
\right).\left(
\begin{array}{c}
 c_{\boldsymbol{k}\sigma} \\
f_{\boldsymbol{k}\sigma} \\
\end{array}
\right) \\\nonumber
&+& \int_{0}^{\beta}d\tau \sum_{\boldsymbol{k}, \sigma} \left(
\begin{array}{cc}
\bar{c}_{\boldsymbol{k}\sigma} & \bar{f}_{\boldsymbol{k}\sigma} \\
\end{array}
\right).\left(
\begin{array}{cc}
 \partial_{\tau} + \epsilon_{\boldsymbol{k}} & V \\
 V & \partial_{\tau} + \epsilon_{f} \\
\end{array}
\right).\left(
\begin{array}{c}
 \delta c_{\boldsymbol{k}\sigma} \\
\delta f_{\boldsymbol{k}\sigma} \\
\end{array}
\right) \\\nonumber
&+& \frac{g^2}{2}\sum_{q}\sum_{k,\sigma}\sum_{p,s}\int_{0}^{\beta}d\tau_1\int_{0}^{\beta}d\tau_2\Big[ \bar{c}_{k,\sigma}(\tau_1)\delta c_{k-q,\sigma}(\tau_1)D^{0}(\tau_1 - \tau_2)\bar{c}_{p-q,s}(\tau_2)c_{p,s}(\tau_2)\\\nonumber
&+& \delta \bar{c}_{k,\sigma}(\tau_1)c_{k-q,\sigma}(\tau_1)D^{0}(\tau_1 - \tau_2)\bar{c}_{p-q,s}(\tau_2)c_{p,s}(\tau_2)\\\nonumber
&+& \bar{c}_{k,\sigma}(\tau_1)c_{k-q,\sigma}(\tau_1)D^{0}(\tau_1 - \tau_2)\bar{c}_{p-q,s}(\tau_2)\delta c_{p,s}(\tau_2) \\\nonumber
&+& \bar{c}_{k,\sigma}(\tau_1)c_{k-q,\sigma}(\tau_1)D^{0}(\tau_1 - \tau_2)\delta \bar{c}_{p-q,s}(\tau_2)c_{p,s}(\tau_2)\Big] \\\nonumber
&-& U\sum_{pqk}\int_{0}^{\beta}d\tau \Big[ \delta\bar{f}_{p, \uparrow}\bar{f}_{q,\downarrow}f_{k,\uparrow}f_{p+q-k, \downarrow} + \bar{f}_{p,\uparrow}\delta\bar{f}_{q,\downarrow}f_{k,\uparrow}f_{p+q-k, \downarrow} \\\nonumber
&+& \bar{f}_{p,\uparrow}\bar{f}_{q,\downarrow}f_{k,\uparrow}\delta f_{p+q-k,\downarrow} + \bar{f}_{p,\uparrow}\bar{f}_{q,\downarrow}\delta f_{k,\uparrow}f_{p+q-k, \downarrow}\Big]
\end{eqnarray}
From Eq. [\ref{pam_Schwinger_Dyson_equation}], we can obtain two independent equations, which are 
\begin{align}
&\int\mathcal{D}[\bar{c}, c]\mathcal{D}[\bar{f}, f] e^{-S_{eff}}\Big[ \delta\bar{c}_{k,\sigma}(\tau) - \delta S_{eff}\bar{c}_{k,\sigma}(\tau)\Big] = 0 \\
&\int\mathcal{D}[\bar{c}, c]\mathcal{D}[\bar{f}, f] e^{-S_{eff}}\Big[\delta\bar{f}_{k,\sigma}(\tau) - \delta S_{eff}\bar{f}_{k,\sigma}(\tau)\Big] = 0
\end{align}
For the first equation, we can set $\delta c = 0, \delta f = 0, \delta\bar{f} = 0$ and obtain
\begin{eqnarray}
&&\delta\bar{c}_{k,\sigma}(\tau) - \delta S_{eff}\bar{c}_{k,\sigma}(\tau)\\\nonumber
&=& \int_{0}^{\beta}d\tau^{\prime}\sum_{k^{\prime}, \sigma^{\prime}} \delta\bar{c}_{k^{\prime},\sigma^{\prime}}(\tau^{\prime})\Bigg[\delta_{k,k^{\prime}}\delta_{\sigma,\sigma^{\prime}}\delta(\tau - \tau^{\prime}) - \Big((\partial_{\tau^{\prime}} + \epsilon_{k^{\prime}})c_{k^{\prime},\sigma^{\prime}}(\tau^{\prime})\bar{c}_{k,\sigma}(\tau) + Vf_{k^{\prime}, \sigma^{\prime}}(\tau^{\prime})\bar{c}_{k,\sigma}(\tau)\Big) \\\nonumber
&-& g^2\sum_{q}\sum_{p,s} \int_{0}^{\beta}d\tau_2\Big[ c_{k^{\prime}-q,\sigma^{\prime}}(\tau^{\prime})\bar{c}_{k,\sigma}(\tau)D^{0}(\tau^{\prime} - \tau_2)\bar{c}_{p,s}(\tau_2)c_{p+q,s}(\tau_2)\Big]\Bigg]
\end{eqnarray}
Plug this into the path integral, we have the equation 
\begin{eqnarray}
\label{holstein_energy}
&& -\Big\langle g^2\sum_{q}\sum_{p,s}  \int_{0}^{\beta}d\tau_2\Big[ c_{k^{\prime}-q,\sigma^{\prime}}(\tau^{\prime})\bar{c}_{k,\sigma}(\tau)D^{0}(\tau^{\prime} - \tau_2)\bar{c}_{p,s}(\tau_2)c_{p+q,s}(\tau_2)\Big]\Big\rangle \\\nonumber
&=& -\delta_{k,k^{\prime}}\delta_{\sigma, \sigma^{\prime}}\delta(\tau^{\prime} - \tau) - \Big[(\partial_{\tau^{\prime}} + \epsilon_{k^{\prime}})G^{cc}_{k,\sigma;k^{\prime},\sigma^{\prime}}(\tau^{\prime} - \tau) + VG^{fc}_{k,\sigma;k^{\prime},\sigma^{\prime}}(\tau^{\prime} - \tau)\Big]
\end{eqnarray}
For the second equation, we can set $\delta f = 0, \delta c = 0, \delta\bar{c} = 0$, and then we have 
\begin{eqnarray}
&&\delta\bar{f}_{k,\sigma}(\tau) - \delta S_{eff}\bar{f}_{k,\sigma}(\tau)\\\nonumber
&=& \sum_{k^{\prime},\sigma^{\prime}}\int_{0}^{\beta}d\tau^{\prime}\delta\bar{f}_{k^{\prime},\sigma^{\prime}}(\tau^{\prime})\Bigg[\delta_{k,k^{\prime}}\delta_{\sigma,\sigma^{\prime}}\delta(\tau - \tau^{\prime}) - \Big(Vc_{k^{\prime},\sigma^{\prime}}(\tau^{\prime})\bar{f}_{k,\sigma}(\tau) + (\partial_{\tau^{\prime}} + \epsilon_f)f_{k^{\prime},\sigma^{\prime}}(\tau^{\prime})\bar{f}_{k,\sigma}(\tau)\Big)\\\nonumber
&+& U\sum_{pq}\Big( \delta_{\sigma^{\prime},\uparrow}\bar{f}_{k,\sigma}(\tau)\bar{f}_{q,\downarrow}f_{p,\uparrow}f_{k^{\prime} + q-p, \downarrow} + \delta_{\sigma^{\prime}, \downarrow}\bar{f}_{p,\uparrow}\bar{f}_{k,\sigma}(\tau)f_{q,\uparrow}f_{p+k^{\prime}-q, \downarrow}\Big)\Bigg]
\end{eqnarray}
Plug this into the path integral, we have 
\begin{eqnarray}
\label{hubbard_energy}
&& \Big\langle U\sum_{pq}\Big( \delta_{\sigma^{\prime},\uparrow}\bar{f}_{k,\sigma}(\tau)\bar{f}_{q,\downarrow}f_{p,\uparrow}f_{k^{\prime} + q-p, \downarrow} + \delta_{\sigma^{\prime}, \downarrow}\bar{f}_{p,\uparrow}\bar{f}_{k,\sigma}(\tau)f_{q,\uparrow}f_{p+k^{\prime}-q, \downarrow}\Big) \Big\rangle \\\nonumber
&=& -\delta_{k,k^{\prime}}\delta_{\sigma, \sigma^{\prime}}\delta(\tau^{\prime} - \tau) -\Big[ VG^{cf}_{k,\sigma;k^{\prime}, \sigma^{\prime}}(\tau^{\prime} - \tau) + (\partial_{\tau^{\prime}} + \epsilon_f)G^{ff}_{k,\sigma;k^{\prime},\sigma^{\prime}}(\tau^{\prime} - \tau)\Big]
\end{eqnarray}
Combining Eq. [\ref{holstein_energy}] and [\ref{hubbard_energy}] together, we have this matrix equation ($\phi$ and $\psi$ are just padding elements for the matrix equation, and their explicit forms are of no concern to us): 
\begin{eqnarray}\label{PAM_eom}
&&\left(
\begin{array}{cc}
E^{g}_{k,\sigma;k^{\prime},\sigma^{\prime}}(\tau^{\prime}, \tau) & \phi \\
 \psi & E^{U}_{k,\sigma;k^{\prime},\sigma^{\prime}}(\tau^{\prime}, \tau) \\
\end{array}
\right) \\\nonumber
&=& \left(
\begin{array}{cc}
 -\delta_{k,k^{\prime}}\delta_{\sigma, \sigma^{\prime}}\delta(\tau^{\prime} - \tau) & 0 \\
 0 & -\delta_{k,k^{\prime}}\delta_{\sigma, \sigma^{\prime}}\delta(\tau^{\prime} - \tau) \\
\end{array}
\right) \\\nonumber
&-& \left(
\begin{array}{cc}
 \partial_{\tau^{\prime}} + \epsilon_{k^{\prime}} & V \\
 V & \partial_{\tau^{\prime}} + \epsilon_f \\
\end{array}
\right) \left(
\begin{array}{cc}
 G^{cc}_{k,\sigma;k^{\prime},\sigma^{\prime}}(\tau^{\prime} - \tau) & G^{cf}_{k,\sigma;k^{\prime},\sigma^{\prime}}(\tau^{\prime} - \tau) \\
 G^{fc}_{k,\sigma;k^{\prime},\sigma^{\prime}}(\tau^{\prime} - \tau) & G^{ff}_{k,\sigma;k^{\prime},\sigma^{\prime}}(\tau^{\prime} - \tau) \\
\end{array}
\right)
\end{eqnarray}
Here, in the above equation, we have defined two quantities 
\begin{eqnarray}\nonumber
&& E^{g}_{k,\sigma;k^{\prime},\sigma^{\prime}}(\tau^{\prime}, \tau) = -\Big\langle g^2\sum_{q}\sum_{p,s}  \int_{0}^{\beta}d\tau_2\Big[ c_{k^{\prime}-q,\sigma^{\prime}}(\tau^{\prime})\bar{c}_{k,\sigma}(\tau)D^{0}(\tau^{\prime} - \tau_2)\bar{c}_{p,s}(\tau_2)c_{p+q,s}(\tau_2)\Big]\Big\rangle \\\nonumber
&& E^{U}_{k,\sigma;k^{\prime},\sigma^{\prime}}(\tau^{\prime}, \tau) = \Big\langle U\sum_{pq}\Big( \delta_{\sigma^{\prime},\uparrow}\bar{f}_{k,\sigma}(\tau)\bar{f}_{q,\downarrow}f_{p,\uparrow}f_{k^{\prime} + q-p, \downarrow} + \delta_{\sigma^{\prime}, \downarrow}\bar{f}_{p,\uparrow}\bar{f}_{k,\sigma}(\tau)f_{q,\uparrow}f_{p+k^{\prime}-q, \downarrow}\Big) \Big\rangle
\end{eqnarray}
If we set $k = k^{\prime}, \sigma = \sigma^{\prime}, \tau = 0, \tau^{\prime} = 0^{-}$ in Equation [\ref{PAM_eom}], then we have 
\begin{eqnarray}
\left(
\begin{array}{cc}
E^{g}_{k,\sigma} & \phi \\
 \psi & E^{U}_{k,\sigma} \\
\end{array}
\right) =  \frac{1}{\beta}\sum_{i\omega_n}\Sigma_{k,\sigma}(i\omega_n)G_{k,\sigma}(i\omega_n)e^{i\omega_n 0^{+}} 
\end{eqnarray}
Therefore, the total interaction energy is 
\begin{eqnarray}
E_{V} &=& \sum_{k,\sigma}\text{Tr} \left(
\begin{array}{cc}
E^{g}_{k,\sigma} & \phi \\
 \psi & E^{U}_{k,\sigma} \\
\end{array}
\right) \\\nonumber
&=& \frac{1}{\beta}\sum_{i\omega_n}\sum_{k,\sigma}\text{Tr}\Big(\Sigma_{k,\sigma}(i\omega_n)G_{k,\sigma}(i\omega_n)\Big)e^{i\omega_n 0^{+}} 
\end{eqnarray}
The kinetic energy is trivial to calculate, which is 
\begin{align}
E_{K} =  \frac{1}{\beta} \text{Tr}\sum_{\boldsymbol{k},\sigma,i\omega_n} \left(
\begin{array}{cc}
 \epsilon_{\boldsymbol{k}}  & V \\
 V & \epsilon_{f}  \\
\end{array}
\right) \left(
\begin{array}{cc}
 G^{cc}_{\boldsymbol{k},\sigma}(i\omega_n) &  G^{cf}_{\boldsymbol{k},\sigma}(i\omega_n) \\
  G^{fc}_{\boldsymbol{k}, \sigma}(i\omega_n) &  G^{ff}_{\boldsymbol{k}, \sigma}(i\omega_n) \\
\end{array}
\right) e^{i\omega_n 0^{+}}
\end{align}
\end{widetext}

Combination of the kinetic energy and the interaction energy gives us the total energy of the system. 

\section{Conclusion}
\label{conclusion}
In this paper, we have demonstrated how to derive the Green function's equation of motion without resort to an explicit Hamiltonian. The equation of motion for Green function is straightforward to get when we have Hamiltonian formulation of the system. However, Hamiltonian formulation of a system is not always available. One such case where we are bereft of an explicit Hamiltonian formulation is when we integrate out phonons to obtain a retarded electron-electron interaction in Holstein model or its generalizations. We have shown that for a simplified toy Holstein model, Green function's equation of motion can be derived via three distinct yet equivalent methods, one of which is to integrate out the phonons and use Schwinger-Dyson equation method. We further show that for periodic Anderson model with electron-phonon interactions, we can equally well derive the equation of motion via Schwinger-Dyson equation method. Our ultimate goal is to calculate the total energy of the system, a goal that can be accomplished with some mathematical manipulations once we have already obtained the equation of motion for electronics Green functions. We showed that the well known formula for calculating the total energy of a system is still applicable here, although for our case that formula should be derived through the Schwinger-Dyson equation method. 

\appendix
\section{Integrate out phonons in periodic Anderson model with electron-phonon interaction}
In this appendix, we will show how to integrate out the phonons in Periodic Anderson model with electron-phonon interactions.
Once the phonons are integrated out, we are left with an effective action that only involves the electrons. In the new effective action, electrons experience a retarded attractive potential. We can interpret this as one electron is attracting another electron by emitting or absorbing a virtual phonon. The advantage of integrating out phonons is that we are left with an effective system that involves only electrons, an effective system that can be conveniently simulated using the continuous time quantum Monte Carlo (CT-QMC) method. We have been using CT-QMC to simulate the effective system that results from integrating out phonons, and it is the study of this system that motivates us to develop the Schwinger-Dyson equation method to derive the total energy formula as shown in this paper. 

The partition function for our original Hamiltonian is 
\begin{align}
Z  = \int\mathcal{D}[\bar{c}, c] \mathcal{D}[\bar{f}, f] \mathcal{D}[\bar{a}, a] e^{-S}
\end{align}
Here, $S$ is the action which is 
\begin{widetext}
\begin{align}
S &= \int_{0}^{\beta}d\tau\sum_{\boldsymbol{k}, \sigma} \left(
\begin{array}{cc}
 \bar{c}_{\boldsymbol{k}\sigma} & \bar{f}_{\boldsymbol{k}\sigma} \\
\end{array}
\right).\left(
\begin{array}{cc}
 \partial_{\tau} + \epsilon_{\boldsymbol{k}} & V \\
 V & \partial_{\tau} + \epsilon_{f} \\
\end{array}
\right).\left(
\begin{array}{c}
 c_{\boldsymbol{k}\sigma} \\
f_{\boldsymbol{k}\sigma} \\
\end{array}
\right) \\\nonumber
&+ \int_{0}^{\beta}d\tau\sum_{\boldsymbol{k}}\bar{a}_{\boldsymbol{k}}(\partial_{\tau} + \omega_0)a_{\boldsymbol{k}} +  \int_{0}^{\beta}d\tau\sum_{\boldsymbol{k}\boldsymbol{q}\sigma} g\bar{c}_{\boldsymbol{k} + \boldsymbol{q}, \sigma}
c_{\boldsymbol{k}\sigma}(a_{\boldsymbol{q}} + \bar{a}_{-\boldsymbol{q}})\\\nonumber
&- U\int_{0}^{\beta}d\tau \sum_{\boldsymbol{k}\boldsymbol{p}\boldsymbol{q}} \bar{f}_{\boldsymbol{p}\uparrow}\bar{f}_{\boldsymbol{q}\downarrow}f_{\boldsymbol{k}\uparrow}f_{\boldsymbol{p} + \boldsymbol{q} - \boldsymbol{k}, \downarrow}
\end{align}
The phononic contribution to the action is  
\begin{eqnarray}
S_{\phi} = \int_{0}^{\beta}d\tau\sum_{\boldsymbol{k}}\bar{a}_{\boldsymbol{k}}(\partial_{\tau} + \omega_0)a_{\boldsymbol{k}} +  \int_{0}^{\beta}d\tau\sum_{\boldsymbol{k}\boldsymbol{q}\sigma} g\bar{c}_{\boldsymbol{k} + \boldsymbol{q}, \sigma}
c_{\boldsymbol{k}\sigma}(a_{\boldsymbol{q}} + \bar{a}_{-\boldsymbol{q}})
\end{eqnarray}

Integration with respect to phonons yields 
\begin{eqnarray}
Z_{eff} &=& \int \mathcal{D}[\bar{a}, a] e^{-S_{\phi}}\\\nonumber
&=& Z_{0} \exp\Bigg(-\sum_{\boldsymbol{q}, m}\bar{J}_{\boldsymbol{q}, m} \frac{1}{i\nu_{m} - \omega_0} J_{\boldsymbol{q}, m}\Bigg)
\end{eqnarray}
Here, the partition function for bare phonons is 
\begin{eqnarray}
Z_{0} = \prod_{\boldsymbol{q}}\frac{1}{1 - e^{-\beta\omega_0}}, 
\end{eqnarray}
and the current $J_{\boldsymbol{q}, m}$ is defined as 
\begin{eqnarray}
J_{\boldsymbol{q},m} &=& \beta^{-1/2}g\sum_{\boldsymbol{k},\sigma}\sum_{i\omega_n}\bar{c}_{\boldsymbol{k} - \boldsymbol{q},\sigma}(i\omega_n - i\nu_m)c_{\boldsymbol{k},\sigma}(i\omega_n)\\\nonumber
&=& g\beta^{-1/2}\int_{0}^{\beta}d\tau e^{i\nu_m\tau} \sum_{\boldsymbol{k},\sigma}\bar{c}_{\boldsymbol{k} - \boldsymbol{q}, \sigma}(\tau)c_{\boldsymbol{k},\sigma}(\tau)
\end{eqnarray}
Its conjugate is 
\begin{eqnarray}
\bar{J}_{\boldsymbol{q}, m} = g\beta^{-1/2}\int_{0}^{\beta}d\tau e^{-i\nu_m\tau}\sum_{\boldsymbol{k}\sigma}\bar{c}_{\boldsymbol{k}\sigma}(\tau)c_{\boldsymbol{k} - \boldsymbol{q}}(\tau) 
\end{eqnarray}
After integrating out the phonons, we have an effective action which is 
\begin{eqnarray}
S^{\phi}_{eff} = \sum_{\boldsymbol{q},m}\bar{J}_{\boldsymbol{q},m}\frac{1}{i\nu_m - \omega_0}J_{\boldsymbol{q}, m} 
\end{eqnarray}

An important property of $J_{\boldsymbol{q}, m}$ is that $J_{-\boldsymbol{q}, -m} = \bar{J}_{\boldsymbol{q}, m}$. Thus, the effective action can be rewritten as 
\begin{eqnarray}
S^{\phi}_{eff} = \frac{1}{2} \sum_{\boldsymbol{q}, m} \bar{J}_{\boldsymbol{q}, m} \frac{-2\omega_0}{\omega_0^2 - (i\nu_m)^2} J_{\boldsymbol{q}, m}
\end{eqnarray}
Fourier transforming back to $\tau$ space, we have 
\begin{eqnarray}
S^{\phi}_{eff} = \frac{1}{2}g^2 \sum_{\boldsymbol{q}}\int_{0}^{\beta}d\tau_1\int_{0}^{\beta}d\tau_2\sum_{\boldsymbol{k}, \sigma}\bar{c}_{\boldsymbol{k},\sigma}(\tau_1)c_{\boldsymbol{k} - \boldsymbol{q}, \sigma}(\tau_1)D^{0}(\tau_1 - \tau_2)\sum_{\boldsymbol{p},s}\bar{c}_{\boldsymbol{p} - \boldsymbol{q}, s}(\tau_2)c_{\boldsymbol{p},s}(\tau_2)
\end{eqnarray}
Here, we have introduced the bare phonon propagator in the effective action, 
\begin{eqnarray}
D^{0}(\tau_1 - \tau_2) &=& \frac{1}{\beta}\sum_{i\nu_m}\frac{2\omega_0}{(i\nu_m)^2 - \omega_0^2}e^{-i\nu_{m}(\tau_1 - \tau_2)}\\\nonumber
&=& -\frac{1}{1-e^{-\beta\omega_0}}(e^{-\omega_0|\tau_1 - \tau_2|} + e^{-(\beta - \vert \tau_1 - \tau_2 \vert)\omega_0})
\end{eqnarray}
The effective action $S^{\phi}_{eff}$ represents the time-retarded density-density interaction between two $c$ electrons mediated by the exchange of a virtual phonon. This interaction differs from the instantaneous Hubbard interaction not only in that it is time-retarded, but also in that the spins of the two electrons do not have to be opposite, as required for on-site Hubbard interactions. 

With this observation, the original partition function can be recast into the form
\begin{eqnarray}
Z = Z_{\text{Bare\_phonons}} \int \mathcal{D}[\bar{c}, c]\mathcal{D}[\bar{f}, f] e^{-S_{eff}}, 
\end{eqnarray}
where, the effective action is 
\begin{align}
S_{eff} &= \int_{0}^{\beta}d\tau\sum_{\boldsymbol{k}, \sigma} \left(
\begin{array}{cc}
 \bar{c}_{\boldsymbol{k}\sigma} & \bar{f}_{\boldsymbol{k}\sigma} \\
\end{array}
\right).\left(
\begin{array}{cc}
 \partial_{\tau} + \epsilon_{\boldsymbol{k}} & V \\
 V & \partial_{\tau} + \epsilon_{f} \\
\end{array}
\right).\left(
\begin{array}{c}
 c_{\boldsymbol{k}\sigma} \\
f_{\boldsymbol{k}\sigma} \\
\end{array}
\right) \\\nonumber
&+ \frac{1}{2}g^2 \sum_{\boldsymbol{q}}\int_{0}^{\beta}d\tau_1\int_{0}^{\beta}d\tau_2\sum_{\boldsymbol{k}, \sigma}\bar{c}_{\boldsymbol{k},\sigma}(\tau_1)c_{\boldsymbol{k} - \boldsymbol{q}, \sigma}(\tau_1)D^{0}(\tau_1 - \tau_2)\sum_{\boldsymbol{p},s}\bar{c}_{\boldsymbol{p} - \boldsymbol{q}, s}(\tau_2)c_{\boldsymbol{p},s}(\tau_2) \\\nonumber
&- U\int_{0}^{\beta}d\tau \sum_{\boldsymbol{k}\boldsymbol{p}\boldsymbol{q}} \bar{f}_{\boldsymbol{p}\uparrow}\bar{f}_{\boldsymbol{q}\downarrow}f_{\boldsymbol{k}\uparrow}f_{\boldsymbol{p} + \boldsymbol{q} - \boldsymbol{k}, \downarrow}
\end{align}
\end{widetext}

In our Monte Carlo simulation, we start with the effective action, and obtain Green functions and self-energies for retarded conduction electrons and instantaneous $f$ electrons. Now that we are bereft of a Hamiltonian formulation of the system, we cannot directly use the formula $E_{V} = \frac{1}{\beta} \sum_{\boldsymbol{k}, \sigma, i\omega_n} \Sigma_{\boldsymbol{k, \sigma}}(i\omega_n)G_{\boldsymbol{k}, \sigma}(i\omega_n)$ which is derived based on the assumption that the system can be described by a Hamiltonian. Here, we propose an alternative method to derive the total energy with the effective action as our starting point. The derivation of the total energy demands the knowledge of the equation of motion for the Green functions. Schwinger-Dyson equation was originally designed to obtain the equation of motion for the Green function based on the path integral principle, and thus can be readily used here to derive the total energy for our system. In the next section, we will give a brief review of the Schwinger-Dyson equation.

\bibliography{ref}
\bibliographystyle{unsrt}

\end{document}